\documentclass[11pt]{article}

\usepackage{amsmath, graphics, graphicx, latexsym, amssymb, amsthm, amsfonts}
\usepackage[mathscr]{eucal}
\usepackage{epsfig}
\usepackage[english]{babel}
\usepackage[square]{natbib}
\usepackage{setspace, enumerate}
\usepackage[pdfauthor={AG}]{hyperref}
\newtheorem{theo}{{\bf{Theorem}}}[section]

\newtheorem{lem}[theo]{{\bf Lemma}}

\newtheorem{rem}{{\bf Remark}}[section]

\newtheorem{defi}{{\bf Definition}}[section]

\renewcommand{\proof}{\noindent{\bf Proof.\ }}
\newcommand{\no}{\nonumber}
\newcommand{\noi}{\noindent}

\newcommand{\la}{\lambda}
\newcommand{\La}{\Lambda}
\newcommand{\calF}{\mathcal{F}}
\newcommand{\calX}{\mathcal{X}}

\newcommand{\vf}{\varphi}

\newcommand{\vp}{\varepsilon}

\catcode`\@=11
\catcode`\@=12

\begin{document}

\title{Convergence of Estimated Option Price in a Regime switching Market}
\author{Anindya Goswami\thanks{IISER, Pune 411008, India; email: anindya@iiserpune.ac.in}\qquad\qquad Sanket Nandan \thanks{IISER, Pune 411008 India; email: sanketnandan@gmail.com}\\ }

%\date{}

\maketitle

{\bf Abstract:}
In an observed semi-Markov regime, estimation of transition rate of regime switching leads towards calculation of locally risk minimizing option price. Despite the uniform convergence of estimated step function of transition rate, to meet the existence of classical solution of the modified price equation, the estimator is approximated in the class of smooth functions and furthermore, the convergence is established. Later, the existence of the solution of the modified price equation is verified and the point-wise convergence of such approximation of option price is proved to answer the tractability of its application in Finance. To demonstrate the consistency in result a numerical experiment has been reported.

{\bf Keywords:} semi-Markov processes, Volterra integral equation, non-local parabolic PDE, locally risk minimizing pricing, optimal hedging

{\bf Classification No.:} 60K15, 91B30, 91G20, 91G60.
\section{Introduction}
Among the researchers of financial market modelling and derivative pricing, regime switching economy is a popular choice, owing to its simplicity and mathematical tractability, along with its ability to incorporate fluctuations of market parameters. See \cite{DKR,MR,RR,DES,BAS,AGMKG} and references therein for more details. Broadly speaking, regime switching economy refers to a class of mathematical models of financial securities where the market parameters such as expected growth rate, volatility, interest rate etc are assumed to evolve as a stochastic process with finite state space. Therefore, for obvious reason, a fair price of a contingent claim under such market assumption, depends on the law of this finite state process. In practice, the law of such regime switching is not known a priori. If the regimes are observed, then one can perhaps estimate the transition rate of the regime switching and calculate an option price using the estimator. This approach would be satisfactory provided the calculated price is `close' to the theoretical price in some sense. In this paper we examine validity of such approach. Since the market, under discussion, is incomplete, there are multiple fair prices. We consider the unique locally risk minimizing price as in \cite{S} and often do not mention explicitly.

\noi In this paper we consider semi-Markov regimes which subsumes the popular Markov regimes as well. It is known that unlike Markov case, the transition rate in this generality cannot be written as a constant matrix. The rate of semi-Markov process turns out to be a matrix valued measurable function on $[0,\infty)$. Nevertheless, it is possible to construct a sequence of MLE of certain finite dimensional projections of the transition rate function and then establish the almost sure convergence of the sequence to the transition rate function \cite{ST, OL}. On the other hand, it appears that the price function of a European option satisfies a differential equation (see \cite{AGMKG} for more details) in which the transition rate appears as a parameter. It is shown in \cite{AJP} that this equation admits a classical solution if the transition rate satisfies certain conditions including differentiability. Therefore it is important to ensure two things. Firstly, if the transition rate function is replaced by a smoothed version of a member of the sequence of MLE, in the differential equation, does the equation still admit a classical solution? Secondly, if one gets a sequence of approximated price in the above manner, does it converge to the theoretical price? We answer both of these in the affirmative. We also illustrate this convergence result by one numerical experiment.

\noi The rest of this paper is arranged as follows. In Section 2 the mathematical model of the financial market and the pricing equation is briefly presented. A particular smooth approximation of the transition rate function is constructed in Section 3. It also contains the convergence result of the smooth approximation. In Section 4, we show that using the sequence of estimator of transition rate function, one can construct an approximating sequence of the European call option price which converges to the true price. Finally some numerical experiments are reported in Section 5.

\section{Market model and pricing} Here we consider a market consisting of only two assets among which one is a locally risk-free asset, such as a money market account and another is a risky asset, such as stock. Let $\{X_t\}_{t \ge 0}$ be a semi-Markov process on a finite state space $\mathcal{X}=\{1,2,\ldots,\theta\}$ with transition probabilities from state $i$ to state $j$ as $p_{ij}$ and conditional holding time distributions given the present state $i$ and next state $j$ as $F(\cdot| i)$. Assume that $(\Omega,\calF,P)$ is the underlying complete probability space which also contains a Wiener process $\{W_t\}_{t \ge 0}$ that is independent to $\{X_t\}_{t \ge 0}$. We model the hypothetical state (assumed to be observable) of the market by $\{X_t\}_{t \ge 0}$ and assume that the bank interest rate, volatility coefficient, and the growth rate evolve as functions of $X_t$ and are denoted as $r(X_t)$, $\sigma(X_t)$ and $\mu(X_t)$ respectively. We denote the consecutive transition times of $X_t$ by  $\{T_n\}_{n \ge 0}$ with $T_0=0$ and the holding time at $t$ by $Y_t$. The price of risk-free and risky assets are denoted by $\{B_t\}_{t \ge 0}$ and $\{S_t\}_{t \ge 0}$ respectively and are given by
\begin{eqnarray}\label{eq4}
B_t &=& \exp(\int_0^t r(X_u) du), \\
\label{eq5} dS_t &=& S_t (\mu(X_t)dt + \sigma(X_t)dW_t) , S_0>0
\end{eqnarray}
where $\sigma$ is positive valued and $r$, $\mu$ are nonnegative maps. Let $\calF_t$ be the filtration of $\calF$ generated by $X_t$ and $S_t$ and satisfying the usual hypothesis. In \cite{AGMKG} it is shown that this market model does admit an equivalent martingale measure, thus it is arbitrage free under \emph{admissible strategy}. It turns out that the above market is incomplete. Therefore, no-arbitrage price of a derivative might not be unique. In this paper we consider locally risk minimizing option price which exists uniquely \cite{AGMKG}. For more details about this pricing approach in a more general setup we refer \cite{S} and the references therein.

\noi In order to obtain a representation of the option price we further assume that
\begin{itemize}
  \item [{\bf{(A1)}}] (i) $F(y|i) \in \mathcal{C}^2([0,\infty))$ $\forall i$\\ %and $f(y|i):=\frac{d}{dy}F(y|i)$ is bounded,\\
(ii) $F(y|i)<1~\forall i$ and $y>0$\\ %$f(y|i)>0~\forall i$ \& $y>0$.
(iii)$(p_{ij})$ is an irreducible probability matrix.
\end{itemize}
We denote the instantaneous transition rate function as $\lambda_{ij}(y)$ which is given by
$$\lim_{\Delta y \downarrow 0} \frac{P(X_{T_{n+1}}=j, y < Y_{T_{n+1}} \leq y+\Delta y | X_{T_{n}}=i, Y_{T_{n+1}} > y)}{\Delta y}.$$
\noi Thus $\lambda_{ij}(y)=p_{ij}\frac{f(y|i)}{1-F(y|i)}$.

\noi In this paper, as an example, we consider a European call option, on ${S_t}$ with strike price $K$ and maturity time $T$. Then the contingent claim $(S_T - K)^{+}$ is $\mathcal{F}_{T}$ measurable. It is shown in \cite{AGMKG} and \cite{AJP} that the locally risk minimizing price of this claim at time $t$ is a function of $t, S_t, X_t$ and $Y_t$ and that, $\vf$ say, is the unique solution in the class of functions of at most linear growth of the initial boundary value problem
\begin{eqnarray}\label{eq1}
\no& &\frac{\partial}{\partial t}\vf(t,s,i,y) +  \frac{\partial}{\partial y}\vf(t,s,i,y) + r(i) s\frac{\partial}{\partial s}\vf(t,s,i,y) + \frac{1}{2} \sigma^2(i) s^2 \frac{\partial^2}{\partial s^2}\vf(t,s,i,y)\\
& & \sum_{j \neq i} \la_{ij}(y) [\vf(t,s,j,0) - \vf(t,s,i,y)] = r(i) \vf(t,s,i,y),
\end{eqnarray}
defined on
\begin{equation}\label{eq2}
\mathcal{D}:=\{(t,s,i,y) \in (0,T) \times \mathbb{R}^+ \times \mathcal{X} \times (0,T) | y \in (0,t)\},
\end{equation}
with boundary conditions
\begin{eqnarray}\label{eq3}
\no& & \lim_{s\downarrow 0}\vf(t,s,i,y)=0, ~~\forall t \in [0,T],\\
& &\vf(T,s,i,y) = (s-K)^+; ~~ s \in \mathbb{R}^+; ~~ 0 \le y \le T; ~~ i=1,2, ... ,\theta.
\end{eqnarray}
In \cite{AJP} it is shown that the above problem is equivalent to a Volterra integral equation of second kind which can be comfortably solved numerically using a step by step quadrature method.

\section{Approximation of transition rate}
We augment the semi-Markov process $X_t$ with the holding time process $Y_t$ to obtain a process $(X,Y)=(X_t,Y_t)_{t \geq 0}$, which is clearly Markov.

\noi Because of the deterministic nature of $(X_t,Y_t)$ during every interval $(T_n,T_{n+1})$, the dynamics of $(X_t,Y_t)$ can be described by a discrete time Markov process $(X_n,Y_n)_{n \ge 0}$ where we define $X_n:=X_{T_n}$ and $Y_n:=Y_{T_n-}=T_n-T_{n-1}$. Thus the augmented Markov process can uniquely be specified with initial distribution as $P(X_{0}=j):=p(j)$, $Y_{T_0}=0$ and the following semi-Markov kernel,

\begin{equation}
P(X_{n+1}=j,Y_{n+1} \leq y | X_0,X_1,...,X_n,Y_1,...,Y_n) := p_{X_n j}F(y|X_n) ~~~~ (a.s.)
\end{equation}
for all $y \in \overline{\mathbb{R}}^{+}$ and $1 \leq j \leq \theta$. We denote $Q_{ij}(y) := p_{ij} F(y|i)~ \forall i \neq j$,
$\lambda_{i}(y):= \sum_{j \in \mathcal{X}, j \neq i} \lambda_{ij}(y)$, $\Lambda_i(y):= \int_0^y \lambda_{i}(u) du$.
We see, $F(y|i)= \sum_{j=1}^{\theta} Q_{ij}(y)$ and also
\begin{eqnarray*}
 \frac{d F(u|i)}{du} = f(u|i) &=& \sum_{j \in \mathcal{X}, j \neq i} p_{ij} f(u|i) = (1-F(u|i)) \sum_{j \in \mathcal{X}, j \neq i} \lambda_{ij}(u).
\end{eqnarray*}
By solving the above ODE for $F$, one obtains 
$$\ln (1-F(y|i))= -\Lambda_i(y), \textrm{ which implies } F(y|i)= 1 - \exp(- \Lambda_i(y)).$$
Hence
\begin{eqnarray}\label{Flambda}
\sum_{j=1}^{\theta} Q_{ij}(y) &=& 1 - \exp(-\Lambda_i(y)).
\end{eqnarray}

\noi Consider a history of augmented Markov process censored at fixed time $\tau$,
$$ \mathcal{H}(\tau) = (X_0, X_1,\cdots, X_{N_\tau}, Y_1, Y_2,\cdots, Y_{N_\tau}, U_\tau),$$
where $N_\tau$ is the number of transitions before time $\tau$ and $U_\tau := \tau - T_{N_\tau}$ is the backward recurrence time. The associated log-likelihood function is maximized to obtain the maximum likelihood estimator(MLE) of the transition rate function, $\lambda_{ij}(\cdot)$.
The likelihood function for $\mathcal{H}(\tau)$ is $$L(\tau) = p(X_0) (1 - \sum_{l=1}^{\theta} Q_{X_{N_\tau}l} (U_\tau)) \prod_{l=0}^{N_\tau-1} p_{X_l X_{l+1}} f(Y_{l+1}|X_l).$$
Thus from \eqref{Flambda},
$$p(X_0)^{-1} L(\tau) = exp(-\Lambda_{X_{N_\tau}}(U_\tau)) \prod_{l=0}^{N_\tau-1} exp(-\Lambda_{X_l}(Y_{l+1})) \lambda_{X_l,X_{l+1}}(Y_{l+1}).$$
\noi Then we consider log-likelihood as
 $$l(\tau) := \log \{p(X_0)^{-1} L(\tau)\} = \sum_{l=0}^{N_\tau-1} (\log \lambda_{X_l,X_{l+1}}(Y_{l+1}) - \Lambda_{X_l}(Y_{l+1}))
 - \Lambda_{X_{N_\tau}}(U_\tau).$$
\noi We consider $(v_k)_{0 \leq k \leq M-1}$, a regular subdivision of $[0,\tau]$ with step $\Delta_\tau = \frac{\tau}{M}$ and $M=\lfloor\tau^{1+\alpha}\rfloor$ where $\alpha >0$, to define for $i\neq j\in \calX$ and $y>0$
\begin{equation}\label{lam}
\lambda_{ij}^*(y) := \sum_{k=0}^{M-1} \lambda_{ijk} 1_{(v_k,v_{k+1}]}(y),
\end{equation}
where $\lambda_{ijk}=\lambda_{ij}(v_k)$. After replacing $\lambda$ by the step function $\lambda^{*}$ in the expression of $l(\tau)$ one obtains,
\begin{equation}\label{modl}
\sum_{i,j \in S} \sum_{k=0}^{M-1} (d_{ijk} \log \lambda_{ijk} - \lambda_{ijk} v_{ik}),
\end{equation}
where
$$v_{ik} := \sum_{l=0}^{N_\tau-1} (Y_{l+1}\wedge v_{k+1} - v_k) 1_{\{ i\}\times(v_k,\infty)}(X_l,Y_{l+1}) + (U_\tau \wedge v_{k+1} - v_k) 1_{\{i\}\times(v_k,\infty)}(X_{N_\tau}, U_\tau),$$
and
$$d_{ijk} := \sum_{l=0}^{N_\tau-1} 1_{\{i\}\times\{j\}\times{(v_k,v_{k+1}]}}(X_l,X_{l+1},Y_{l+1}).$$
\noi Hence the estimator of $\lambda_{ijk}$ which maximizes the functional in \eqref{modl}, is given by
\begin{equation}
\hat\lambda_{ijk} =
\begin{cases}
\ d_{ijk}/v_{ik} ~~\textrm{if}~~ v_{ik} > 0;\\
\ 0 ~~~~~~~~~~~\textrm{otherwise.}
\end{cases}
\end{equation}
\noi Thus we obtain an estimator of $\lambda^{*}_{ij}(y)$, given by,
\begin{equation}\label{hatl}
\hat\lambda_{ij}(y,\tau) = \sum_{k=0}^{M-1} \hat\lambda_{ijk} 1_{(v_k,v_{k+1}]}(y) + \hat\lambda_{ij0} 1_{\{0\}}(y).
\end{equation}

\noi We have the following result.
\begin{lem}\label{1}
Fix $\alpha \in (0,1/2)$. Under the assumptions of (A1), the estimator $\hat\lambda_{ij}(\cdot,\tau)$, is uniformly strongly consistent for $\lambda_{ij}(\cdot)$, on $[0,T]$ in the sense that
$$\max_{i\neq j} \sup_{y \in [0,T]} |\hat\lambda_{ij}(y, \tau) - \lambda_{ij}(y)| \rightarrow 0, ~~\textrm{almost surely, as}~~ \tau \rightarrow \infty.$$
\end{lem}
\proof If (A1) holds, one can directly derive from the definition of $\la_{ij}(y)$ that $\lambda_{ij}(y)=p_{ij}\frac{f(y|i)}{1-F(y|i)}$. Now by summing over $j\in \calX \setminus \{i\}$ and then integrating on $[0,y)$ both the sides, one obtains $\La_i(y)=-\ln (1- F(y |  i))$. Thus under assumption (A1) $\La_i(y)$ is in $\mathcal{C}^2([0,\infty))$. Hence, for any $i\neq j$,  $\lambda_{ij}: [0,\infty) \to [0,\infty)$ is well defined and continuously differentiable. Now the rest of the proof follows from Theorem 1(b) of \cite{OL}.\qed

\noi In view of the above Lemma, we fix $\alpha \in (0,1/2)$ for the rest of this paper. Note that $\hat\lambda(\cdot,\tau)$ is a step function and thus it is discontinuous. We aim to obtain an approximation of $\lambda$ in the class of smooth functions so that the approximation can be used to obtain an approximated price function by solving appropriate system of differential equations. To this end we consider $B^2$-spline interpolation of $\hat\lambda(\cdot,\tau)$. We first extend it on $\mathbb{R}$ by assigning $\hat\lambda(y,\tau)=\hat\lambda(0,\tau)~ \forall y<0$ and $\hat\lambda(y,\tau)=\hat\lambda(T_1,\tau)~ \forall y> T_1:= \emph{arg}\max_{y\in [T,\tau]}\hat\lambda(y,\tau)$ and setting $v_k:=k\Delta_\tau$, for all $k \in \mathbb{Z}$. Set
\begin{eqnarray*}
B_k^2(y)&=&\frac{(y-v_k)^2}{(v_{k+2}-v_k)(v_{k+1}-v_{k})}1_{[v_k,v_{k+1})}(y)\\ &&+ \left( \frac{(y-v_k)(v_{k+2}-y)}{(v_{k+2}-v_{k+1})(v_{k+2}-v_k)} + \frac{(v_{k+3}-y)(y-v_{k+1})}{(v_{k+2}-v_{k+1})(v_{k+3}-v_{k+1})} \right) 1_{[v_{k+1},v_{k+2})}(y) \\
&& + \frac{(v_{k+3}-y)^2}{(v_{k+3}-v_{k+2})(v_{k+3}-v_{k+1})}1_{[v_{k+2},v_{k+3})}(y).
\end{eqnarray*}
\noi For each $i\neq j$, we interpolate the data points $(v_k,\hat\lambda_{ij(k-1)})$ and we denote the spline interpolation of $\hat\lambda_{ij}$ as $\tilde{\lambda}_{ij}$ which is given by
\begin{equation}\label{4}
\tilde\lambda_{ij}(y, \tau):=\sum_{k=-\infty}^{\infty} \hat\lambda_{ij}(v_{k+2}, \tau) B_k^2(y)=\sum_{k=-\infty}^{\infty} \hat\lambda_{ij(k+1)} B_k^2(y).
\end{equation}
Since for each $i\neq j$, $\hat\lambda_{ijk}$ is non-negative for every $k$, from the property of B-spline, $\tilde\lambda_{ij}$ is a non-negative function.
\begin{lem}\label{2}
Under (A1)
$$\max_{i\neq j} \sup_{y \in [0,T]} | \hat\lambda_{ij}(y,\tau) - \tilde\lambda_{ij}(y,\tau)|\to 0$$ as $\tau \to \infty$.
\end{lem}
\proof From Theorem 6.6.4 of \cite{KC}, for each $i\neq j$, $\sup_{y \in [0,T]} | \hat\lambda_{ij}(y,\tau) - \tilde\lambda_{ij}(y,\tau)| \le 2\omega_{[0,T]}(\hat\lambda_{ij}(\cdot,\tau);\Delta_\tau)$ where $\omega_{I}(f,\delta):=\sup\{ |f(t)-f(s)| |  t,s\in I, ~ |t-s|=\delta \}$ is the modulus of continuity. Again it follows from Lemma \ref{1}, that for a given $\vp>0$ there is a $N$ such that $P(N < \infty)=1$ and for $\tau \ge N$, $\max_{i\neq j} \sup_{y \in [0,T]} |\hat\lambda_{ij}(y, \tau) - \lambda_{ij}(y)| <\vp$. Thus for $\tau > N$, $\omega_{[0,T]}(\hat\lambda_{ij} (\cdot,\tau);\Delta_\tau) \le \omega_{[0,T]}(\lambda_{ij}(\cdot);\Delta_\tau)+2\vp$. Since $\lambda_{ij}(\cdot)$ is continuous, its modulus of continuity converges to zero as $\Delta_\tau\to 0$. Therefore, we get as $\tau \to \infty$, $\omega(\hat\lambda_{ij}(\cdot,\tau);\Delta_\tau) \to 0$. Hence the result. \qed

\noi Lemma \ref{1} and \ref{2} lead to the following result.
\begin{theo}\label{5}
Under (A1)
$$\max_{i\neq j} \sup_{[0,T]} |\tilde\lambda_{ij}(y,\tau) - \lambda_{ij}(y)| \rightarrow 0, ~\textrm{a.s., as}~ \tau \rightarrow \infty.$$
\end{theo}

\section{Approximation of price function}

\noi It is shown in the previous section that a convergent sequence of smooth approximations of transition rate function can be constructed using a combination of non-parametric MLE and $B_2$ spline. Having this result, it is tempting to solve \eqref{eq1}-\eqref{eq3} with the smooth approximation of transition rate to obtain an approximation of price function. Needless to mention that such approximation is reliable only when certain continuous dependency of the solution on the transition rate function is established for the concerned initial boundary value problem. Such result is not readily available for the non-local degenerate type of parabolic PDE, we consider here. In this section we establish the convergence of such approximate price function to the true price. For the sake of preciseness we propose the following definition.

\begin{defi}
Let $\tilde{\lambda}:\calX\times \calX\times [0,\infty)\to [0,\infty)$ an approximation of the transition rate function $\lambda$ and $\vf$ the solution of the problem \eqref{eq1}-\eqref{eq3}. If \eqref{eq1}-\eqref{eq3}, after the function $\lambda$ is replaced by $\tilde{\lambda}$, admits a unique classical solution $\tilde{\vf}$, then $\tilde{\vf}$ is called the \emph{TBA (Transition rate based approximation) of $\vf$ with parameter $\tilde{\lambda}$}.
\end{defi}
\noi \cite{AJP} presents a set of fairly general sufficient condition on $\tilde{\lambda}$, which ensures existence of a unique classical solution of the modified equation.
\begin{lem}\label{lem2}
Under (A1), the TBA of $\vf$ (as in \eqref{eq1}-\eqref{eq3}) with parameter $\tilde \lambda(\cdot, \tau)$ (as in \eqref{4})exists for sufficiently large value of $\tau$.
\end{lem}
\proof We define $\tilde \lambda_{i}(y,\tau):=\sum_{j\neq i}\tilde \lambda_{ij}(y, \tau)$, $\tilde\La_i(y,\tau)=\int_0^y\tilde \lambda_{i}(y,\tau) dy $, $\tilde F(y |  i, \tau):=1-\exp(-\tilde\La_i(y,\tau))$, and $\hat p_{ij}:=\int_0^\infty \frac{\tilde\la_{ij}(y,\tau)}{\tilde \lambda_{i}(y,\tau)}1_{(0,\infty)}(\tilde\lambda_{i}(y,\tau)) d\tilde F(y |  i,\tau)$. If we can show that there is a $\tau_0>0$ such that for $\tau\ge \tau_0$, the sufficient conditions in \cite{AJP} hold i.e.,
(i) $\tilde \lambda(\cdot, \tau)$ is in $C^1([0,\infty))$, (ii) $\lim_{y\to \infty}\tilde\La_i(y,\tau)=\infty$ for each $i$, and (iii) $\hat p_{ij}$ is irreducible, then the existence of TBA of $\vf$ follows from Theorem 3.2 of \cite{AJP}.

\noi We note that from the property of B-spline, (i) holds. Again since for each $i$, $\lim_{y\to \infty}\La_i(y)=\lim_{y\to \infty}-\ln (1- F(y |  i))=\infty$, from Theorem \ref{5} there is a $\tau_0>0$ such that for $\tau\ge \tau_0$, (ii) holds. We first note that to prove (iii), it is sufficient to show that if $p_{ij}>0$ for some $i\neq j$, then $\hat p_{ij}$ is also positive. Now if $p_{ij}>0$, $\la_{ij}(y)(=p_{ij}\frac{f(y|i)}{1-F(y|i)})$ is not identically zero. Hence using Theorem \ref{5}, there exists a $\tau_1>0$ such that for any $\tau >\tau_1$, $\tilde \la_{ij}(y,\tau)$ is not identically zero. Thus
\begin{eqnarray*}\label{}
\hat p_{ij}&=& \int_0^\infty \frac{\tilde\la_{ij}(y,\tau)}{\tilde \lambda_{i}(y,\tau)}1_{(0,\infty)}(\tilde\lambda_{i}(y,\tau)) d\tilde F(y | i,\tau)\\
&=&\int_0^\infty \tilde\la_{ij}(y,\tau) e^{-\tilde \Lambda_{i}(y,\tau)} dy\\
&>&0.
\end{eqnarray*}\qed

\noi We state the main result below.
\begin{theo}\label{3}
Under (A1) for a large  value of $\tau$, let, $\tilde\vf^\tau$ be the TBA of $\vf$ with parameter $\tilde \lambda(\cdot, \tau)$. Then, $\tilde\vf^\tau$ converges to $\vf$ point-wise, as $\tau \rightarrow \infty$.
\end{theo}
\noi In order to prove this theorem, we need the the following lemma.
\begin{lem}\label{Lem1}
Let $\{S_t\}_{t\ge 0}$ be as in \eqref{eq5}. Then
\begin{equation}\label{66}
E[\int_0^T  S_t dt]\leq\frac{S_0}{C} \left( e^{C T} - 1 \right)
\end{equation}
for some positive constant $C$.
\end{lem}
\proof
We know, \eqref{eq5} has a closed form solution given by
\begin{equation}
S_t=S_0 \exp\left[\int_0^t \{\mu(X_u)-\frac{1}{2}\sigma^2(X_u)\}du + \int_0^t \sigma(X_u)dW_u\right].
\end{equation}
We introduce the following constants
$$c:=\max_{i \in \mathcal{X}}\left\{\mu(i)-\frac{1}{2}\sigma^2(i)\right\}, ~~~ d:=\max_{i \in \mathcal{X}}\{ \sigma^2(i)\}.$$
Clearly,
\begin{equation}\label{stock}
S_t \le S_0 \exp(c t) \exp\left(\int_0^t \sigma(X_u)dW_u\right).
\end{equation}
We observe,
\begin{eqnarray*}
\int_0^t  \sigma(X_u)dW_u &=& \sum_{n=1}^\infty \int_{T_{n-1}\wedge t }^{T_n \wedge t } \sigma(X_{T_{n-1}})dW_u\\
&=& \sum_{n=1}^\infty \sigma(X_{T_{n-1}}) (W_{T_n \wedge t }-W_{T_{n-1} \wedge t })
\end{eqnarray*}
Hence, the conditional distribution of $\int_0^t  \sigma(X_u)dW_u$ given $\mathcal{F}^X_t $ is normal with mean zero and variance $$\sum_{n=1}^\infty \sigma^2(X_{T_{n-1}}) ({T_{n}\wedge t } - {T_{n-1} \wedge t }),$$ where $\calF_t ^X$ is the filtration of $\calF$ generated by $X=\{X_u\}_{u\in [0,t]}$.

\noi Now using the formula of variance of a log-normal random variable, we get,
\begin{eqnarray*}
E\left[\exp\left(\int_0^t  \sigma(X_u) dW_u\right)\right]
&=&E\left[E\left[\exp\left(\int_0^t  \sigma(X_u) dW_u\right)\Big| \calF_t ^X \right]\right]\\
&=& E\left[\exp\left(\frac{1}{2}\sum_{n=1}^\infty \sigma^2(X_{T_{n-1}}) ({T_{n}\wedge t } - {T_{n-1} \wedge t })\right)\right]\\
&\le& E\left[\exp\left(\frac{d}{2}\sum_{n=1}^\infty ({T_{n}\wedge t } - {T_{n-1} \wedge t })\right)\right]\\
&=& \exp\left(\frac{d}{2} t \right).
\end{eqnarray*}
Using above inequality we obtain from \eqref{stock},
$$E(S_t ) \le S_0 e^{c t } e^{\frac{d}{2} t } = S_0 e^{\left( c+ \frac{d}{2} \right) t }.$$
Since $S_t$ is nonnegative we apply Tonelli's theorem and the above relation to get,
\begin{eqnarray*}
E\left[\int_0^T  S_t  dt \right]&=&\int_0^T  E S_t  dt \\
&\leq& S_0 \int_0^T e^{\left( c+ \frac{d}{2} \right) t } dt \\
&=& \frac{S_0}{(c+\frac{d}{2})} \left( e^{\left( c+ \frac{d}{2} \right) T} - 1 \right).
\end{eqnarray*}\qed

\noi {\bf Proof of Theorem \ref{3}.}
We define the difference of the functions $\vf$ and its TBA $\tilde\vf^\tau$ to be,
$$\psi^\tau(t,s,i,y) := \vf(t,s,i,y) - \tilde\vf^\tau(t,s,i,y).$$

\noi Now, by considering the initial boundary value problems satisfied by $\vf(t,s,i,y)$ and $\tilde\vf^\tau(t,s,i,y)$ respectively, one obtains directly that $\psi^\tau$ satisfies the following initial boundary value problem,
\begin{eqnarray*}
& &\frac{\partial}{\partial t}\psi(t,s,i,y) +  r(i)s \frac{\partial}{\partial s}\psi(t,s,i,y) + \frac{1}{2}\sigma^2(i)s^2 \frac{\partial^2}{\partial s^2}\psi(t,s,i,y) + \sum_{j \neq i} \la_{ij}(y) (\psi(t,s,j,0) -  \psi(t,s,i,y)) \\
& & = r(i) \psi(t,s,i,y) - \sum_{j \neq i} (\lambda_{ij}(y)-\tilde\lambda_{ij}(y,\tau))( \tilde\vf^\tau(t,s,j,0) - \tilde\vf^\tau(t,s,i,y)),
\end{eqnarray*}
defined on
$$
\mathcal{D}:=\{(t,s,i,y) \in (0,T) \times \mathbb{R}^+ \times \mathcal{X} \times (0,T) | y \in (0,t)\},
$$
with conditions
 $$\lim_{s\downarrow 0}\psi(t,s,i,y)=0, ~~\forall t \in [0,T],$$
$$\psi(T,s,i,y) =0, ~~ s \in \mathbb{R}^+; ~~ 0 \le y \le T; ~~ i=1,2,\cdots,\theta.$$

\noi We rewrite the above system of equations as,
\begin{equation}
\frac{\partial}{\partial t} \psi(t,s,i,y) + \mathcal{L} \psi(t,s,i,y) = r(i) \psi(t,s,i,y) - f^\tau(t,s,i,y),
\end{equation}
where
$$(\mathcal{L}\psi)(t,s,i,y):= \left[r(i)s \frac{\partial}{\partial s} + \frac{1}{2}\sigma^2(i)s^2 \frac{\partial^2}{\partial s^2}\right] \psi(t,s,i,y) + \sum_{j \neq i} \la_{ij}(y) (\psi(t,s,j,0) -  \psi(t,s,i,y)),$$
and
$$f^\tau(t,s,i,y):=\sum_{j \neq i} (\lambda_{ij}(y)-\tilde\lambda_{ij}(y,\tau))( \tilde\vf^\tau(t,s,j,0) - \tilde\vf^\tau(t,s,i,y)).$$
Note that, $\mathcal{L}$ is the infinitesimal generator of $(\tilde S_t,X_t,Y_t)$ satisfying
$$d\tilde S_t = \tilde S_t(r(X_{t-})dt + \sigma(X_{t-})dW_t),$$
where $X_t$ is a semi-Markov process with transition rate $\lambda_{ij}(y)$ and $Y_t$ is the holding time process. Then, using Feynman-Kac formula,
\begin{equation}
\psi^\tau(t,s,i,y) = E[\int_t^T \exp\left(-\int_t^{t'} r(X_u)du\right) \eta^\tau (t')dt' | \tilde S_t=s, X_t=i, Y_t=y],
\end{equation}
where, $$\eta^\tau(t)=f^\tau(t,\tilde S_t,X_t,Y_t).$$

\noi From Theorem \ref{5}, we have the uniform convergence of $\tilde\lambda_{ij}(y,\tau)$ to $\lambda_{ij}(y)$ in $y \in [0,T]$, almost surely. Or, in other words we get, for any given $\epsilon (>0)$, $\exists N$ s.t. $P(N < \infty)=1$ and for $\tau \ge N,$
\begin{equation}\label{ch}
|\tilde\lambda_{ij}(y,\tau)-\lambda_{ij}(y)|<\vp~~~ \forall y \in [0,T].
\end{equation}
\noi Using Theorem 3.2 of \cite{AJP}, there exists constants $k_1$ and $k_2$ such that $0 \le \tilde\vf^\tau(t,s,i,y) \le k_1+k_2 s$ for all $i,y$ and $\tau\ge N$. Hence from \eqref{ch},
$$f^\tau(t,s,i,y)\le \vp(k_1+k_2 s).$$
Therefore, by applying Lemma \ref{Lem1}, we assert that for all large $\tau$, $|\eta^\tau(t,\omega)|$ is dominated by a fixed integrable function.
Thus using dominated convergence theorem, we can see
\begin{eqnarray*}
\lim_{\tau} \psi^\tau(t,s,i,y) &=& \lim_\tau E[\int_t^T \exp\left(-\int_t^{t'} r(X_u)du\right) \eta^\tau(t') dt' | \tilde S_t=s, X_t=i, Y_t=y]\\
&=&  E[\int_t^T \exp\left(-\int_t^{t'} r(X_u)du\right) \lim_\tau \eta^\tau (t')dt' | \tilde S_t=s, X_t=i, Y_t=y]\\
&=& 0.
\end{eqnarray*}
Hence, $\tilde\vf^\tau$ converges to $\vf$ point-wise, as $\tau \rightarrow \infty$.\qed
\begin{rem}
In this section, as an example, we consider a European call option to illustrate the the convergence of TBA of price. It is important to note that one can have similar convergence results for other derivatives such as put option price, barrier option price, compound option price etc. as the prices satisfy similar system of PDEs.
\end{rem}

\section{Numerical experiment}
We firstly illustrate the convergence result as in Lemma \ref{1} through considering an example of a semi-Markov process and estimating its transition rate over different time scales of the history of the process. In order to compute and present the actual estimation error, we simulate the example having three hypothetical states \{$1,2,3$\} and the holding time distribution as $$f(y):={y e^{-y}} ~~ \textrm{for}~ y > 0,$$ and we choose the transition matrix as
 $$ (p_{ij})=
 \left( \begin{array}{ccc}
0.0 & 0.1 & 0.9\\
0.4 & 0.0 & 0.6 \\
0.7 & 0.3 & 0.0 \end{array} \right) .$$
Therefore, the theoretical transition rate function for any $y \ge 0$, is given by,
$$\lambda_{ij}(y)=p_{ij}\frac{f(y|i)}{1-F(y|i)}=p_{ij}\frac{y}{y+1} ~~~\forall i,j \in S ~\textrm{and}~ i \neq j .$$

\begin{figure}[h]
\includegraphics[width=0.8\textwidth]{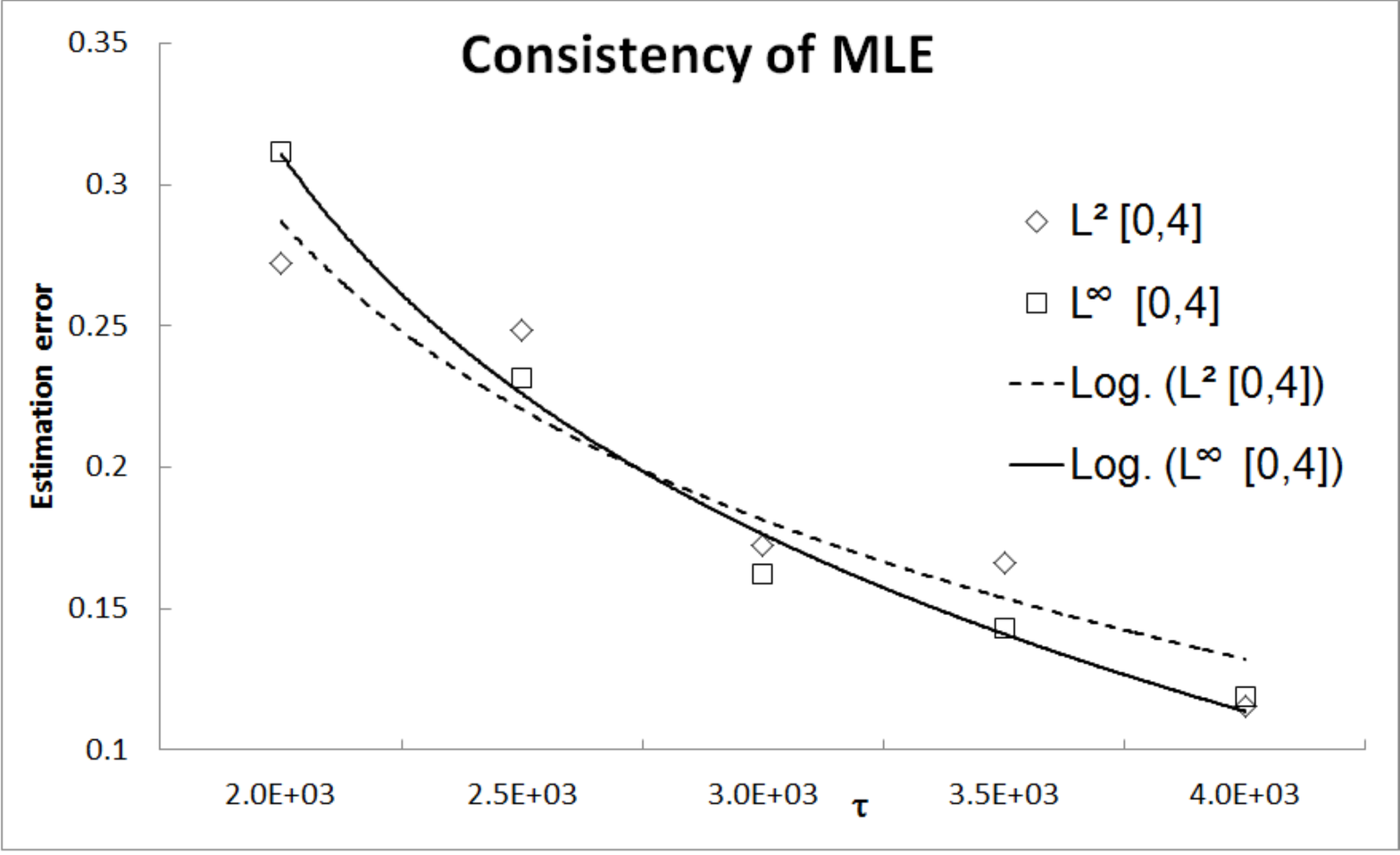}
  \caption{Convergence of MLE}\label{fig1}
\end{figure}

\begin{figure}[h]
\includegraphics[width=0.8\textwidth]{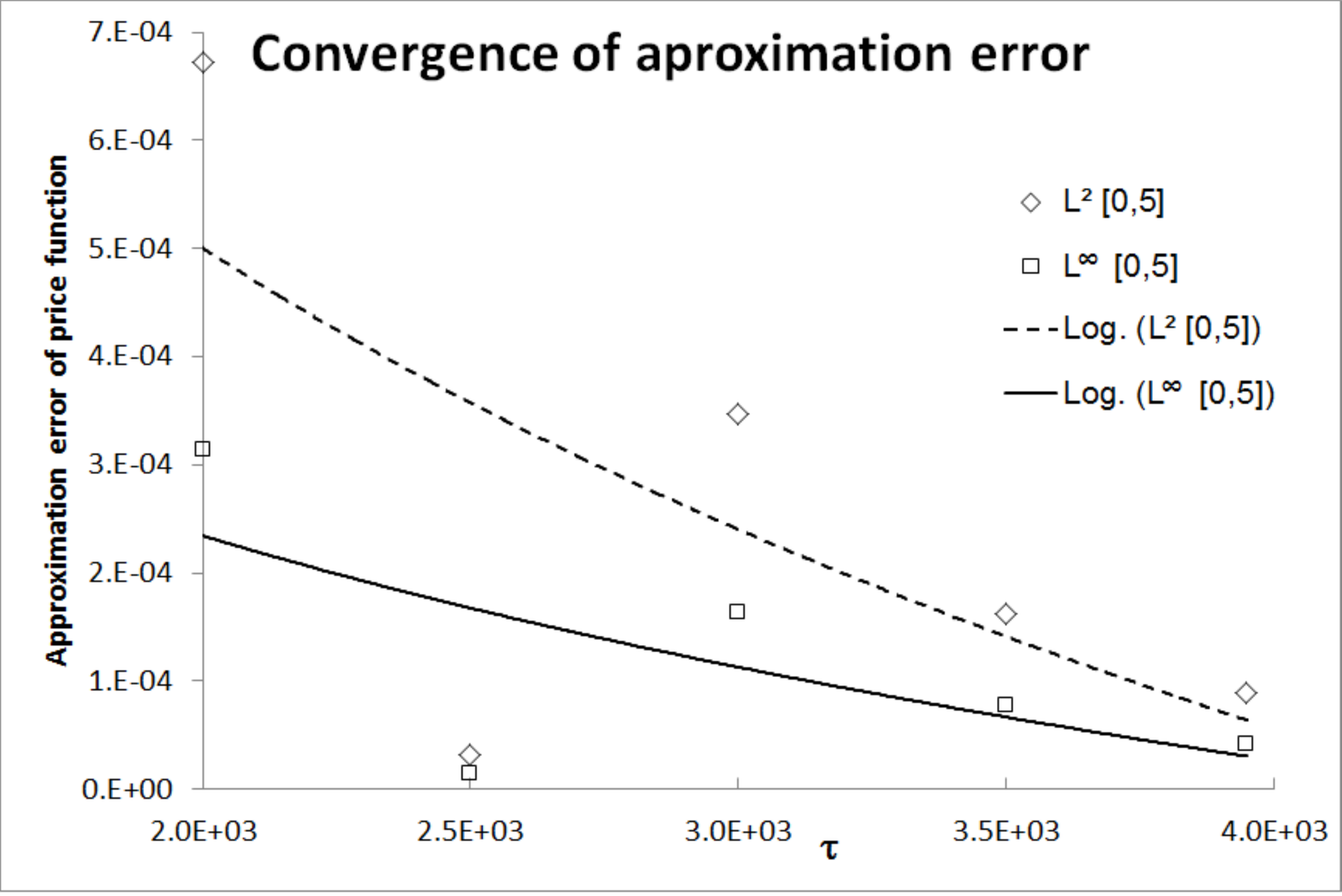}
  \caption{Convergence of approximation error}\label{fig2}
\end{figure}

\noi The convergence as $\tau \rightarrow \infty$ is illustrated by plotting two different norms of the difference between $\hat\la$ and $\la^*$ on $[0,T]$ where $T=4$ for many different values of $\tau$. Although the convergence result is established only for sup norm, but we find it interesting to illustrate the $L^2$ norm also.
In the Figure \ref{fig1},  $\tau$ variable is taken on the horizontal axis and the computed norms of error is plotted on the vertical axis. In the plots, logarithmic trend lines are fitted to visualize the trends.
\noi To illustrate the convergence of TBA of a European call option price function we consider a market where the volatility and instantaneous interest rate at each regime are given as follows
$$(r(1),r(2),r(3))=(0.3,0.6,0.7)$$
$$(\sigma(1),\sigma(2),\sigma(3))=(0.2,0.2,0.2).$$
The strike price $K$ is 1 and the maturity $T$ is 1 unit. Figure \ref{fig2} shows convergence of TBA of price function through the plot of sup norm of the error for the estimation, computed as $$ Error := || \psi^\tau(0,\cdot,i,0) 1_{[0,5]} ||_{sup}. $$ The $L^2$ norm of $\psi^\tau(0,\cdot,i,0) 1_{[0,5]}$ is also plotted and finally a logarithmic trend line is added to each of the data series.

{\bf Acknowledgement:} The authors are grateful to Jayant Deshpande and Mrinal K. Ghosh for very useful discussions.


\begin{thebibliography}{99}

\bibitem[Basak {\it et al} (2011)]{BAS} Basak G. K., Ghosh Mrinal K. and Goswami A., Risk Minimizing Option Pricing for a Class of Exotic Options in a Markov-Modulated Market,  Stoch. Ann. App. 29:2(2011), 259-281.
    \bibliographystyle{plainnat}

\bibitem[Deshpande \& Ghosh (2008)]{DES} Deshpande A. and Ghosh M. K., Risk Minimizing Option Pricing in a Regime Switching Market, Stoch. Ann. App. 26(2008), 313-324.
\bibliographystyle{plainnat}

\bibitem [DiMasi {\it et al} (1994)]{DKR} DiMasi G. B., Kabanov Y. and Runggaldier W. J., Mean-Variance hedging of options on stocks with Markov volatility. Theory Probab. Appl., Vol. 39 (1994), 173-181.
\bibliographystyle{plainnat}

\bibitem[Ghosh \& Goswami (2009)]{AGMKG} Ghosh M. K. and Goswami A., Risk Minimizing Option Pricing in a Semi-Markov Modulated Market, SIAM J. Control Optim. 48(2009), 1519-1541.
\bibliographystyle{plainnat}

\bibitem[Goswami {\it et al} (2015)]{AJP}Goswami, A., Patel, J., Shevgaonkar, P., A system of degenerate non-local parabolic PDE and application, 2015. 	arXiv:1506.01467 [math.AP].
\bibliographystyle{plainnat}


\bibitem[Joberts \& Rogers (2006)]{RR} Joberts A. and Rogers L. C. G., Option pricing with Markov-modulated dynamics, SIAM J. Control Optim. 44(2006), 2063-2078.
\bibliographystyle{plainnat}

%\bibitem[Kallianpur \& Karandikar (2000)]{KK} Kallianpur Gopinath and Karandikar Rajeeva L., Introduction to Option Pricing Theory, Birkhäuser Boston, 2000.
%\bibliographystyle{plainnat}

\bibitem[Kincaid \& Cheney (2010)]{KC} Kincaid D. and Cheney W., Numerical Analysis: Mathematics of Scientific Computing, American Mathematical Society; 3rd Revised edition (2002).
\bibliographystyle{plainnat}

\bibitem[Mamon \& Rodrigo (2005)]{MR} Mamon R. S. and Rodrigo M. R., Explicit solutions to European options in a regime switching economy, Operations Research Letters 33(2005), 581-586.
\bibliographystyle{plainnat}

\bibitem[Ouhbi \& Limnios (1999)]{OL}
Ouhbi, B. Limnios, N., Nonparametric estimation for semi-Markov processes based on its hazard rate functions. Stat. Inference Stoch. Process., 2(1999), 151-173.
\bibliographystyle{plainnat}

\bibitem[Sansom \& Thomson (2001)]{ST} Sansom, John and Thomson, Peter, Fitting hidden semi-Markov models to breakpoint rainfall data, J. Appl. Probab. 38A (2001), 142–157. 
\bibliographystyle{plainnat}

\bibitem[Schweizer (2001)]{S}
Schweizer, M., A guided tour through quadratic hedging approaches(2001). In E.Jouini, J.Cvitani\'c, $\&$ M.Musiela (Eds.), Option pricing interest rates and risk management. Cambridge: Cambridge University Press, 538-574.
\bibliographystyle{plainnat}

\end{thebibliography}
\end{document}